\documentstyle[amsfonts,epsfig,cite,a4wide]{article}
\newcommand{\r}[1]{\ref{#1}}
\newcommand{\lb}[1]{\label{#1}}

\newcommand{\bc}{\begin{center}}
\newcommand{\ec}{\end{center}}
\newcommand{\be}{\begin{equation}}
\newcommand{\ee}{\end{equation}}
\newcommand{\bea}{\begin{eqnarray}}
\newcommand{\eea}{\end{eqnarray}}
\newcommand{\ba}[1]{\begin{array}{#1}}
\newcommand{\ea}{\end{array}}

\newcommand{\bt}[1]{\begin{table}[ht]\centering\begin{tabular}{#1}}
\newcommand{\et}[1]{\end{tabular}\caption{\small#1}\end{table}}

\begin{document}


\thispagestyle{empty}

\begin{flushright}

{\small
{\texttt hep-th/dddmmyy}\\
{\sl February 2006}}

\end{flushright}

\begin{center}

{\large {\bf Why Randall-Sundrum Scenarious are not Compatible with an Orbifolded Fifth Dimension}}

\vspace{2 truecm}

{\bf Pedro Castelo Ferreira}\\[10mm]
CENTRA, Instituto Superior T\'ecnico, Av. Rovisco Pais, 1049-001 Lisboa, Portugal\\[8mm]
Dep. de F\'{\i}s., UBI, Rua Marqu\^es D'\'{A}vila e Bolama, 6200-081 Covilh\~a, Portugal\\

\vspace{2 truecm}
{\bf\sc Abstract}
\begin{minipage}{15cm}
\vspace{2mm}
We show the (already known) fact that Randall-Sundrum scenarious although compatible with a ${\mathbb{Z}}_2$
orbifold symmetry cannot hold regularity of the fields at the orbifold planes in the absence of boundary actions
and respective jumps of the fields. This makes the models mathematical inconsistence and invalidate the inclusion
of such models in a higher dimensional theory such as string theory. For completeness we point out some directions
already in the literature and in progress.
\end{minipage}

\end{center}

\vfill
\begin{flushleft}
Dedicated to a friend: Ian I. Kogan (not to be published)\\
\end{flushleft}

\newpage

The brane world concept was first introduced by Akama~\cite{Akama} in 1982 (see also~\cite{Antoniadis,Visser,Squires}).
Later such scenarious were also considered in~\cite{ovrut} and exploit in a very simplistic manner
by Randall and Sundrum~\cite{RS_0,RS_1} using warpped geometries $M=M_4\times (S_1/{\mathbb{Z}}_2)$ [RSI]
and $M=M_4\times ({\mathbb{R}}/{\mathbb{Z}}_2)$ [RSII].
The geometry used on these approachs is
\bea
ds^2=K(y)(-dt^2+dx^idx_i)+dy^2\ .
\nonumber
\eea
The solution of the eom compatible with the orbifold symmetry ${\mathbb{Z}}_2$ is
\bea
K(y)=\exp\{-k|y-y_0|\}\ .
\nonumber
\eea

For RSI, in order for this solution to be compatible with a compact coordinate it is necessary to sew
together segments as shown in figure~\r{fig.1}.
\begin{figure}[ht]
 \begin{picture}(200,110)(50,-10)
  \put(50,-20){\epsfig{file=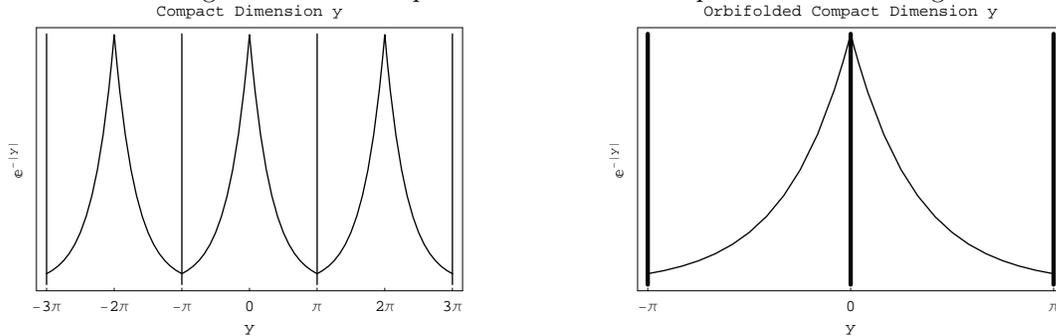,width=160mm}}
 \end{picture}
\caption{RSI solutions on a compact dimension (the thin vertical lines represent the interval identifications)
and respective orbifold (the thick lines represent the orbifold planes.)\lb{fig.1}}
\end{figure}
The main problem is that it is not possible to have a regular field at all the points of a compact coordinate.
Upon orbifolding under ${\mathbb{Z}}_2$ we still have the same problem. At model level the usual
approach is to consider appropriate brane actions localized at the orbifold planes that justify the
jumps on the derivatives.

As for RSII the only solution compatible with ${\mathbb{Z}}_2$ symmetry is $K(y)=\exp\{-k|y|\}$. Again this field is not
regular at $y=0$. Again at model level upon orbifolding and considering appropriate boundary actions we can justify the
jumps on the derivatives.

So the main drawback of RS scenarious is that, at most, can be considered at model level and
cannot be included in any more fundamental higher theory as M/string-theory.

In the context of brane worlds there are already several works in the literature
that give solutions for the problem presented here:
\begin{itemize}
\item Not considering an orbifold symmetry at all~\cite{KG,SW,DDWR}.
\item By considering aditional scalar fields and perturbing the RS solution for $K(y)$ obtaining a regular solution
at the orbifold points~\cite{NM,pospelov}.
\item Considering FRW type of geometries and enlarged gauge groups~\cite{B_1,B_2} that allow for periodic solutions.
\end{itemize}

This last approach further deserves at least to consider generic $N(y)\neq 1$ in order
to understand if there are significative changes in the results obtained. We stress again that a warpped geometry
\textit{a la} RS is not an option.

Work supported by SFRH/BPD/17683/2004.

\end{document}